# Mpemba Effect in Crystallization of Polybutene-1

Jinghua Liu[1], Jingqing Li[1], Binyuan Liu[2], Ian W. Hamley[3], Shichun Jiang*[1]

[1]School of Materials Science and Engineering, Tianjin University, Tianjin, 300072, China

[2]Hebei Key Laboratory of Functional Polymer Materials, School of Chemical Engineering and Science, Hebei University of Technology, Tianjin 300130, China

[3]School of Chemistry, Pharmacy and Food Biosciences, University of Reading, Whiteknights, Reading RG6 6AD, United Kingdom

* Corresponding author: scjiang@tju.edu.cn

**Abstract:** The Mpemba effect and its inverse can be understood as a result of nonequilibrium thermodynamics. In polymers, changes of state are generally non-equilibrium processes. However, the Mpemba effect has been rarely reported in the crystallization of polymers. In the melt, polybutene-1 (PB-1) has the lowest critical cooling rate in polyolefins and tends to maintain its original structure and properties with thermal history. A nascent PB-1 sample was prepared by using metallocene catalysis at low temperature, and the crystallization behavior and crystalline structure of the PB-1 were characterized by DSC and WAXS. Experimentally, a clear Mpemba effect is observed not only in the crystallization of the nascent PB-1 melt in form II but also in form I obtained from the nascent PB-1 at low melting temperature. It is proposed that this is due to the differences in the chain conformational entropy in the lattice which influence conformational relaxation times. The entropy and the relaxation time can be predicted using the Adam-Gibbs equations, whereas non-equilibrium thermodynamics is required to describe the crystallization with the Mpemba effect.

**Keywords:** Mpemba effect, Crystallization of polybutene-1, Dynamics of polymer chain, non-equilibrium thermodynamics

1. Introduction

The Mpemba effect was first systematically observed in freezing water and has recently been reported in various systems, including magnetic alloys, polymers, Clathrate hydrate, ionic liquids, spin glasses, and driven granular gases [1-11]. The Mpemba effect is considered to be a counterintuitive relaxation phenomenon, where an initially hot system cools down faster than an identical system initiated at a warm temperature when both are quenched by a cold bath that is at the same temperature, and the cold bath is large enough relative to either sample so that the temperature of the cold bath does not change. Based on the principle of entropy maximisation in systems that remain at or near thermal equilibrium, the entropy of the initially hot system should be greater than that of the warm one before quenching, and it is impossible to observe the Mpemba effect in a system that passes through all intermediate temperatures during slow cooling. However, when a system is rapidly quenched by placing it in contact with a cold bath, the Mpemba effect can be observed in systems which undergoing a phase transition occasionally. Because of the complexity of the phenomenon, there is still no extensively accepted exact mechanism to understand the Mpemba effect in water and the mechanism or conditions for the occurrence of the Mpemba effect have been controversial. Evaporation [12], convection currents [13], dissolved gases [14], supercooling [15], particular properties of hydrogen bonding [16], freezing point depression by solutes [14], and a difference in the nucleation temperatures of ice nucleation sites between samples [17] have all been proposed to explain the observations. The effect has been under debate, due both to the diversity of proposed explanations and the notable difficulties in replicating the experimental results. In recent years, a theoretical breakthrough and subsequent experimental work have revealed much about the fundamental mechanism [5, 18, 19].

Polymer crystallization is a central topic in polymer physics as it is both of vital importance for the application properties of solid polymeric materials, and of fundamental scientific interest. The understanding of the exact mechanism of crystallization presents a peculiar challenge unique in materials science, which is primarily because crystallization from the melt invariably leads to out-of-equilibrium lamellar morphologies that are hard to model with equilibrium phase thermodynamics [20]. Starting in 1957 [21-23] with the first clear insights into the basic structure of polymeric solids many authors contributed to the development of concepts to describe the crystallization process. In spite of considerable progress in the characterization of the semi-crystalline state of polymers, general agreement on the basic steps followed in the formation of polymer crystallites in a polymer melt has not yet been reached. In particular, after an apparently settled understanding based on the early theories of Hoffman and Lauritzen [24], contradicting experiments were presented about twenty years ago, which indicated a passage through an intermediate mesomorphic phase [25].

Polymer chains consist of a great number of monomers and different side units that form various architectures. The chain structure, including various topologies such as short or long branches, entanglement loops, tie molecules and end groups or tails, is the distinguishing feature of polymer molecules different from small molecules [26]. The various topologies of polymer chains are crucial to the chain dynamics and the chain conformational entropy [27, 28], which lead to the relaxation of polymer chains with the unique property that they can undergo non-equilibrium and even beyond equilibrium processes. Unique properties such as visco-elasticity including enthalpic and entropic elasticity of polymers, is susceptible to ambient temperature. The Williams-Landel-Ferry equation [29] has been widely used to describe the effect of temperature on relaxation, and the Adam-Gibbs equations [30] can be applied to illustrate the temperature dependent relaxation and the conformational entropy. In recent years much work has been done on the equilibrium thermodynamic properties arising from non-equilibrium processes [31-35]. A major breakthrough in this field is the work of Jarzynski [33], which can help to estimate and understand the free energy difference between two states as a suitable average of the work done on polymer systems by forcing the transition in a finite time.

The critical rate of cooling a liquid to a glass can be used to understand the polymer chain dynamics in the melt [36, 37]. Hence, the critical cooling rate of a brisk or fast moving chain is higher than that of a lazy or slow one. For a semi-crystalline polymer, it is the lowest cooling rate that can be used to obtain the 100% amorphous glass state from the melt [37]. The critical cooling rates of the polyolefins, such as high-density polyethylene, isotactic polypropylene, and isotactic polybutene-1 (iPB-1), are >1, 000, 000 K/s, 1000 K/s, and 10 K/s [38-40], respectively. These values for the critical cooling rates of polyolefins, lead to the description of iPB-1 as a strong polyolefin with slow moving chains. iPB-1 is a polymer with polymorphic structures that can form form I, II, III and I′ crystals under different conditions [41]. For iPB-1, the conformation and entropy do not change substantially with temperature, which leads to the anomalous crystallization behavior and formation of the metastable form II [42]. A long-term spontaneous unavoidable transition from form II to form I has become a bottleneck for industrial applications of iPB-1 and is therefore a well-known transition [41]. Here, a metallocene polybutene-1 with 70% tacticity was prepared at 35 °C to obtain nascent samples with less chain symmetry and low melt points of form II and form I. Here we provide clear experimental evidence for the Mpemba effect in the crystallization of the nascent PB-1 samples. The crystallization behavior displays an obvious Mpemba effect in both freshly crystallized form II and form I. These observations provide valuable insights into polymer crystallization and the associated polymer dynamics.

**2. Experimental section**

2.1. Sample preparation

The PB-1 with low isotacticity was synthesized by using metallocene catalysts at 35 °C to obtain nascent samples with less chain symmetry and low melting points of form II and form I. The catalyst has been destroyed by addition of acidified ethanol (10%) after terminal of the synthesis of the sample, then the destroyed catalyst was washed off and the synthesized nascent PB-1 was deposited, filtered and dried. The synthesis was performed in Hebei University of Technology [43], and the mass average molecular weight of the synthesized syndiotactic PB-1 is $2.0 \times 10^4$ g·mol$^{-1}$ and polydispersity is 1.9 measured by Gel Permeation Chromatography (GPC) with high-temperature trichlorobenzene (150 °C) as the mobile phase. The molecular weight is the relative molecular weight converted by the Mark–Houwink equation. The characterization of Nuclear Magnetic Resonance ($^{13}$C NMR) showed that the tacticity of the PB-1 is 69.5%.

2.2. Differential scanning calorimetry (DSC) characterization

A DSC 450 (Linkam Scientific Instruments Ltd) equipped with a LNP95 apparatus as liquid nitrogen controller for cooling was used. Aliquots of (7.0 ± 0.1) mg were weighed and placed in DSC pans. Thermal analyses were executed under a high-purity nitrogen atmosphere to protect the samples from oxidation. The nascent PB-1 samples were first heated to 120 °C, 130 °C, 140 °C and 150 °C and kept for 10 min, and then cooled to room temperature (25 °C) at a cooling rate of 10 °C·min$^{-1}$ to obtain DSC cooling curves during the non-isothermal crystallization. The resulting as-crystallized form II samples were then reheated at 10 °C·min$^{-1}$ to 160 °C to obtain the form II melting curves with the melting temperatures ($T_{m, II}$) determined. The melting peaks of form II were integrated to obtain the corresponding melting enthalpies $\Delta H_{m, II}$ and the crystallinities $X_{c, II}$ of form II were calculated, respectively, according to $X_c = \Delta H_m/\Delta H_m^0$. Here, the melting enthalpy of the ideal crystal of form II is $\Delta H_{m, II}^0 = 62$ J g$^{-1}$ [44]. For further investigation, similar thermal protocols and crystallinity analyses were performed for form II obtained after melting at 120 °C and 150 °C and the corresponding transformed form I. The isothermal crystallization behaviors of the PB-1 samples at 70 °C were carried out after being heated to various temperatures (120 °C, 130 °C, 140 °C and 150 °C) and kept for 10 min or heated to 130 °C and kept for various times (10 min, 30 min, 60 min). The cooling rate from various temperatures (120-150 °C) to 70 °C for the isothermal crystallization was 30 °C/min. The areas under the crystallization peaks were integrated to obtain the evolution of the corresponding relative crystallinity $X_r$ with the isothermal crystallization time $t$. The details of the thermal protocols are shown in Scheme 1 for the related figures.

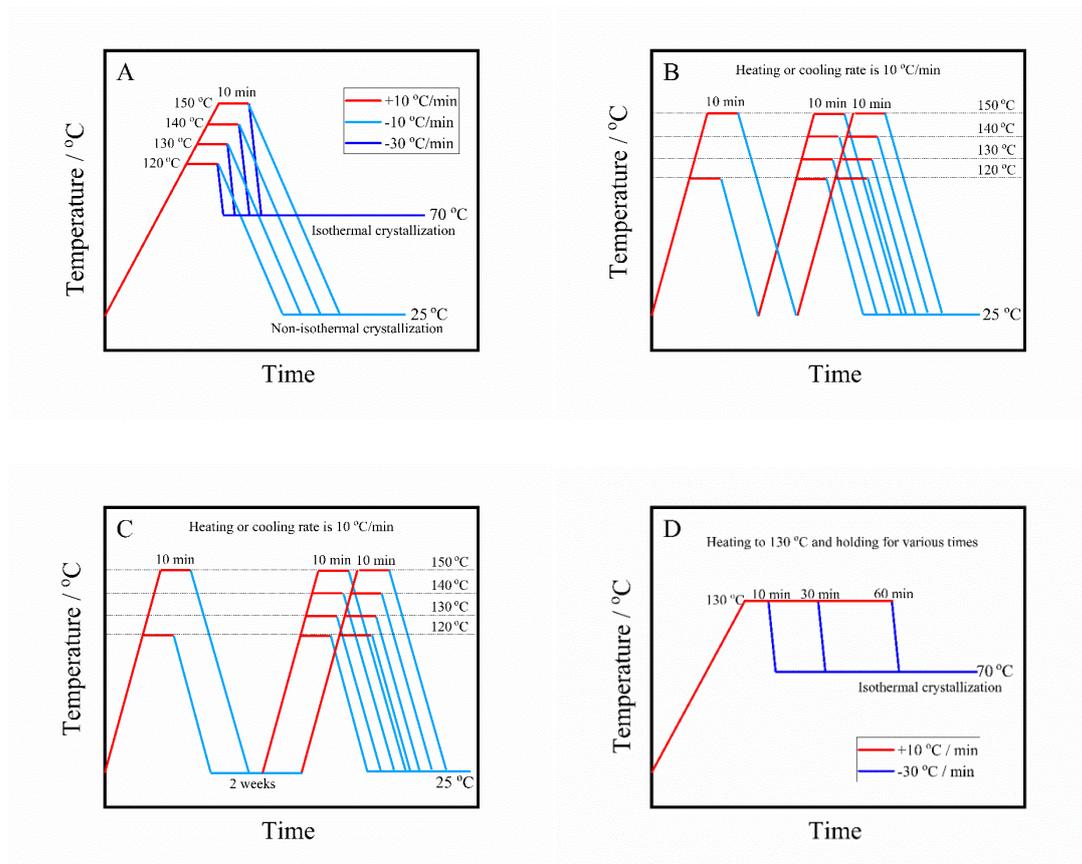

Scheme 1. Illustrations of thermal protocols designed for samples(A: the thermal protocols for the nascent samples; B: the thermal protocols for the form II; C: the thermal protocols for the form I; D: the thermal protocols the samples heating to 130 °C and holding for various times).

2.3. Wide angle X-ray scattering (WAXS) investigation

The WAXS investigations were carried out on beamline 1W2A at Beijing Synchrotron Radiation Facility (BSRF, $\lambda=1.54$ Å) at room temperature. The measurements were carried out using a Rigaku Ultima IV diffractometer equipped with Cu-Kα radiation ($\lambda= 0.154$ nm) and the sample-to-detector distance was 130 mm.

# 3. Results and discussion

The thermal behavior and structure of the nascent PB-1 sample was characterized by DSC and WAXS as shown in the DSC first heating curve (Fig. 1A) and WAXS profile (Fig. 1B). The DSC heating curve displays two melting peaks during heating and the WAXS profile shows the presence of form I' and form III in the nascent PB-1. The peaks in Fig.1B were indexed according to reference 62. During the heating process, form I and form III will melt and recrystallize to form II. Therefore, the two melting peaks in the first heating curve are related to the melting of the form I' and form III (low temperature melting peak), and the melting of the form II (high temperature melting peak [45]), respectively.

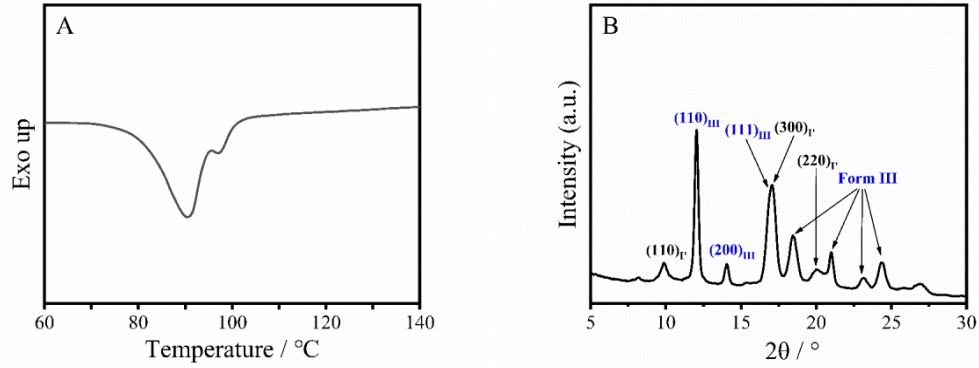

FIG. 1. DSC first heating curve(A) and WAXS profile measured at room temperature (B) of the nascent PB-1.

The synthesized nascent PB-1 samples were first heated to various temperatures and then cooled to room temperature at a cooling rate of 10 °C min$^{-1}$ to observe the crystallization behavior. The obtained DSC cooling curves of the PB-1 are shown in Fig. 2A, which show that the peak crystallization temperatures ($T_c$) shift to higher temperature with initial heating temperature. The $T_c$s obtained are plotted against initial heating temperature in Fig. 2B. The obtained $T_c$s indicate that the supercooling for crystallization of the samples cooled from higher initial heating temperature is lower than that cooled from the lower one. We consider whether this is an indication for an Mpemba effect in crystallization of the nascent PB-1. In order to confirm the phenomenon, the isothermal crystallization behavior of the samples at 70 °C was analysed out after heating to various temperatures. The evolution of the corresponding relative crystallinity $X_r$ with the isothermal crystallization time $t$ is shown in Fig. 2C. The half time for crystallization ($t_{1/2}$) in Fig. 2D was obtained from the results in Fig. 2C, which clearly shows that the $t_{1/2}$ decreases with the initial heating temperature.

The higher the initial temperature, the smaller the half-crystallization time $t_{1/2}$ and the faster is the crystallization rate. In other words, high initial heating temperature causes crystallization to occur at a higher temperature when the cooling rate is constant, which is an obvious evidence for the Mpemba effect during the crystallization of the nascent PB-1.

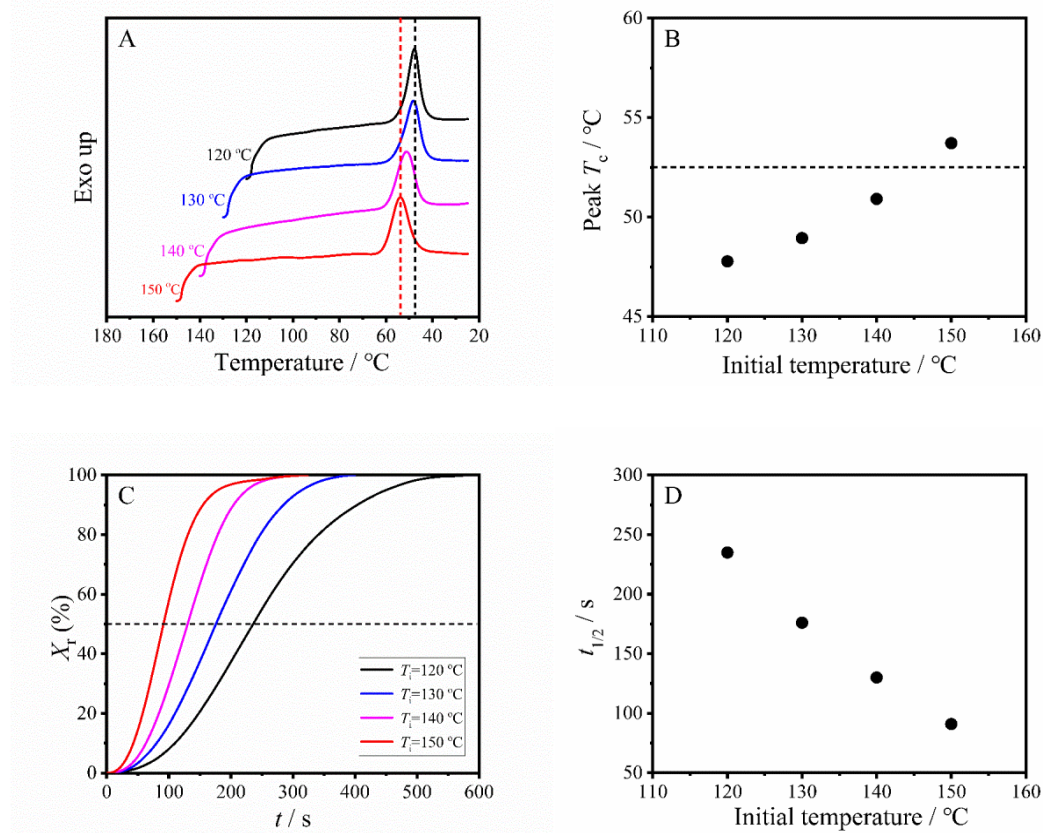

FIG. 2. (A) DSC cooling curves of nascent PB-1 after heating to the indicated temperatures. (B) Initial heating temperature related peak crystallization temperatures of the nascent PB-1 samples. (C) The relative crystallization fraction ($X_r$) of the nascent PB-1 during isothermal crystallization at 70 °C after heating to the indicated temperatures. (D) The $t_{1/2}$ of the isothermal crystallized nascent PB-1 obtained from Fig. 2C. (The thermal protocols are shown in Scheme 1A.)

In this work, the heating process of all nascent PB-1 samples was identical. Therefore, it should be the different initial heating temperatures that lead to the difference in crystallization behaviors. The conformation of a polymer chain adapts and changes depending on the temperature of the state it is passing through [30], the nascent PB-1 samples will produce different amorphous chain conformations when kept for the same time at different temperatures. Due to the long relaxation time of the PB-1 with slow chain motions, the conformation of the chain cannot be adjusted in time when the temperature changes during the cooling process. That is to say, the tendency to maintain the original conformation stabilizes the chain conformational entropy of the chain, which is the origin of the Mpemba effect.

Before cooling, although all initially heated PB-1 chains have amorphous conformations, the chains heated to lower temperatures would be more ordered with lower conformational entropy than those heated to higher temperatures. In general, the ordered or pre-ordered chain conformations can be used as nucleation sites or mesophases to facilitate subsequent crystallization, as in memory effects or self-seeding crystallization of polymers [46-49]. However, the Mpemba effect found in the nascent PB-1 suggests that the more ordered chain conformation formed at lower temperatures cannot serve as nucleation sites or intermediate mesophases to facilitate the crystallization of the PB-1 melt at $T_c = 70\ ^oC$.

Is the Mpemba effect found in the nascent PB-1 related to the crystal form or to the initial heating temperatures? It is known that the equilibrium melting temperatures of crystal form I and crystal form II are 128 °C and 141 °C respectively[50]. Thus, the nascent PB-1 samples were heated to 120 $^oC$ (below the equilibrium melting temperature of PB-1 form II) and 150 $^oC$ (above the equilibrium melting temperature of PB-1 form I) for 10 min and then cooled to 25 $^oC$ to obtain the form II. The thermal properties and crystalline structure of the samples obtained by heating of the nascent PB-1 were confirmed by DSC heating curves (Fig. 3A) and WAXS profiles (Fig. 3B). The DSC heating curves indicate that there is only melting of form II in both samples obtained, however, there is a little more form I' but less form II in the sample obtained by heating to 120 ℃ as shown by the WAXS profiles. The melting point of the sample obtained by heating to 150 ℃ is higher than that obtained by heating to 120 ℃.

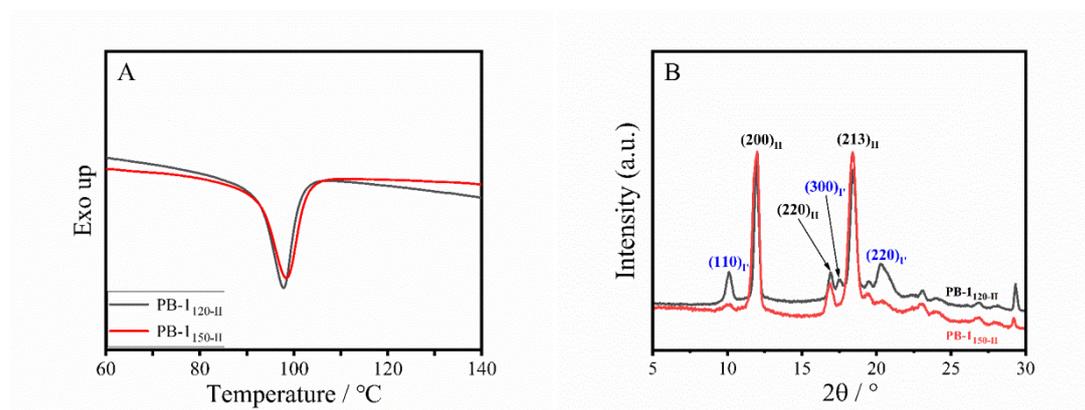

FIG. 3. DSC heating curve (A) and WAXS profile (B) of form II samples of PB-1 formed after heating to 120 °C and 150 °C. (The thermal protocols are shown in Scheme 1B.)

The obtained form II samples were heated to various initial heating temperatures for 10 min and then cooled to 25 °C with DSC characterization, as shown in Fig. 4. The peak $T_c$s obtained from the DSC cooling curves are shown in Fig. 5. The peak $T_c$s of the form II obtained from the nascent PB-1 after heating to 120 °C increase with the initial heating temperature during cooling process. However, the peak $T_c$s of the form II obtained from the nascent PB-1 after heating to 150 °C decrease with the initial heating temperature. The results in Fig. 5 indicate that the supercooling for the form II samples obtained by heating to 120 °C decreases with the initial heating temperature and increases with the initial heating temperature for the form II samples obtained by heating to 150 °C. Therefore, there is an Mpemba effect in the former case but not the latter. The Mpemba effect observed in the PB-1 form II melting is obviously dependent on the thermal history related to the formation of PB-1 form II.

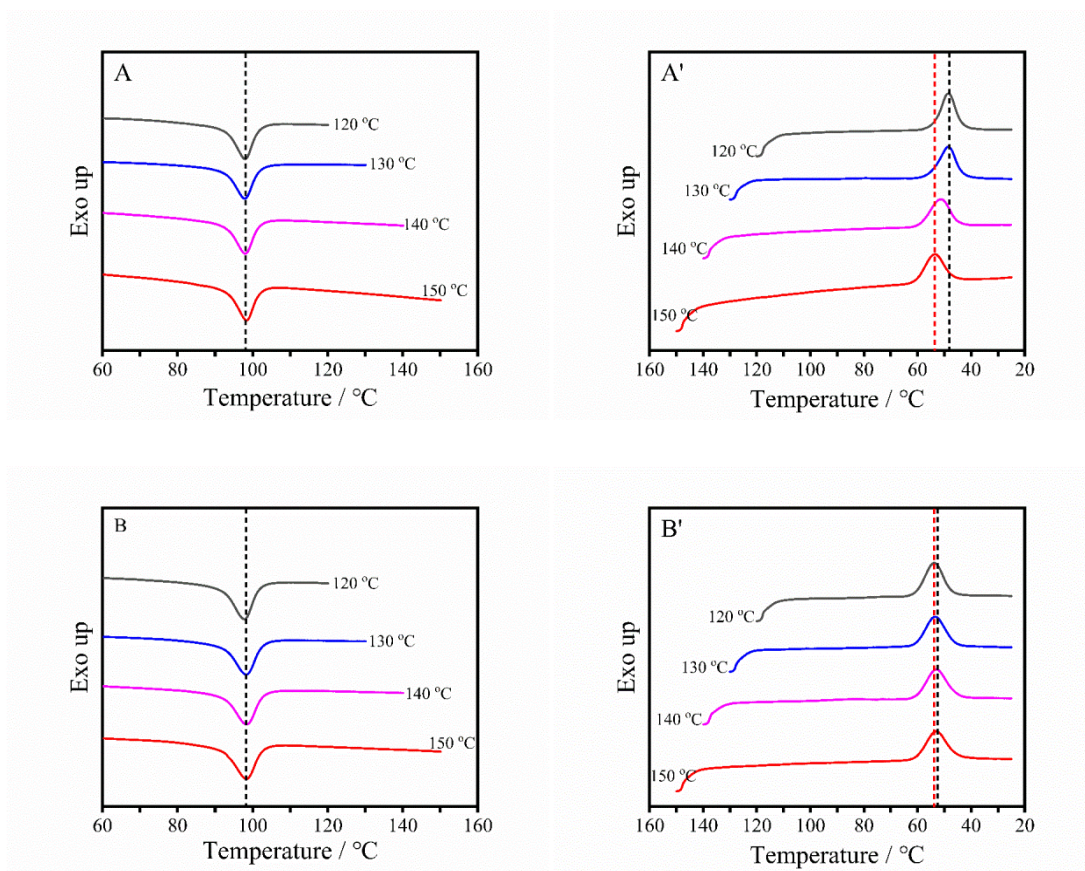

FIG. 4. Initial heating temperature related heating and cooling curves of the form II samples of PB-1 obtained by heating to 120 °C (A and A′) and 150 °C (B and B′). (The thermal protocols are shown in Scheme 1B.)

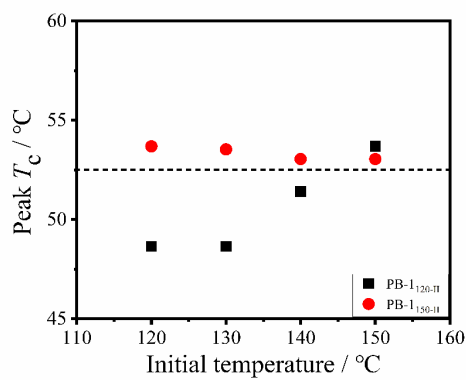

FIG. 5. Initial heating temperature related peak $T_c$s of form II samples of PB-1 formed after heating to 120 °C and 150 °C. (The thermal protocols are shown in Scheme 1B.)

We further investigated whether the Mpemba effect could be observed in the melt crystallization of the form I that transformed from the form II obtained by heating to low or high temperatures. The PB-1 form II samples obtained by heating to 120 °C and 150 °C completely transformed into form I after annealing 14 days at room temperature. The thermal properties and crystalline structure of the transformed samples were characterized by DSC and WAXS as shown in the DSC heating curves (Fig. 6A) and WAXS profiles (Fig. 6B). The results in Fig. 6 show that only form I is present in the transformed PB-1 samples.

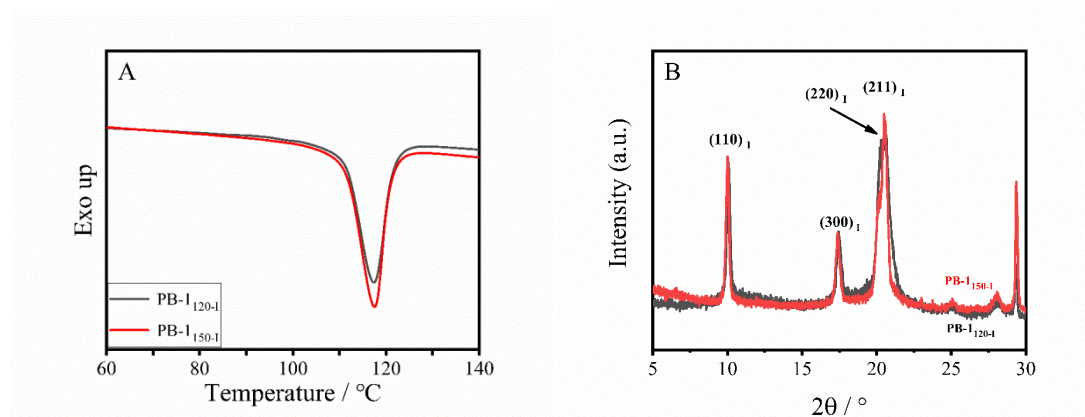

FIG. 6. DSC heating curve (A) and WAXS profile (B) of form I samples of PB-1 formed after heating to 120 °C and 150 °C. (The thermal protocols are shown in Scheme 1C.)

The transformed form I PB-1 samples were heated to various initial temperatures and then cooled to room temperature, and the obtained DSC heating and cooling curves are shown in Fig. 7. As shown in Fig. 8, the $T_c$s of the transformed form I obtained from the DSC cooling curves (Fib. 7A′ and 7B′) indicate that there is a Mpemba effect in the crystallization of the form I samples originally obtained by heating to 120 °C, while the form I samples originally obtained by heating to 150 °C crystallize normally, which is similar to the melting crystallization behavior of the form II samples obtained directly from the nascent PB-1.

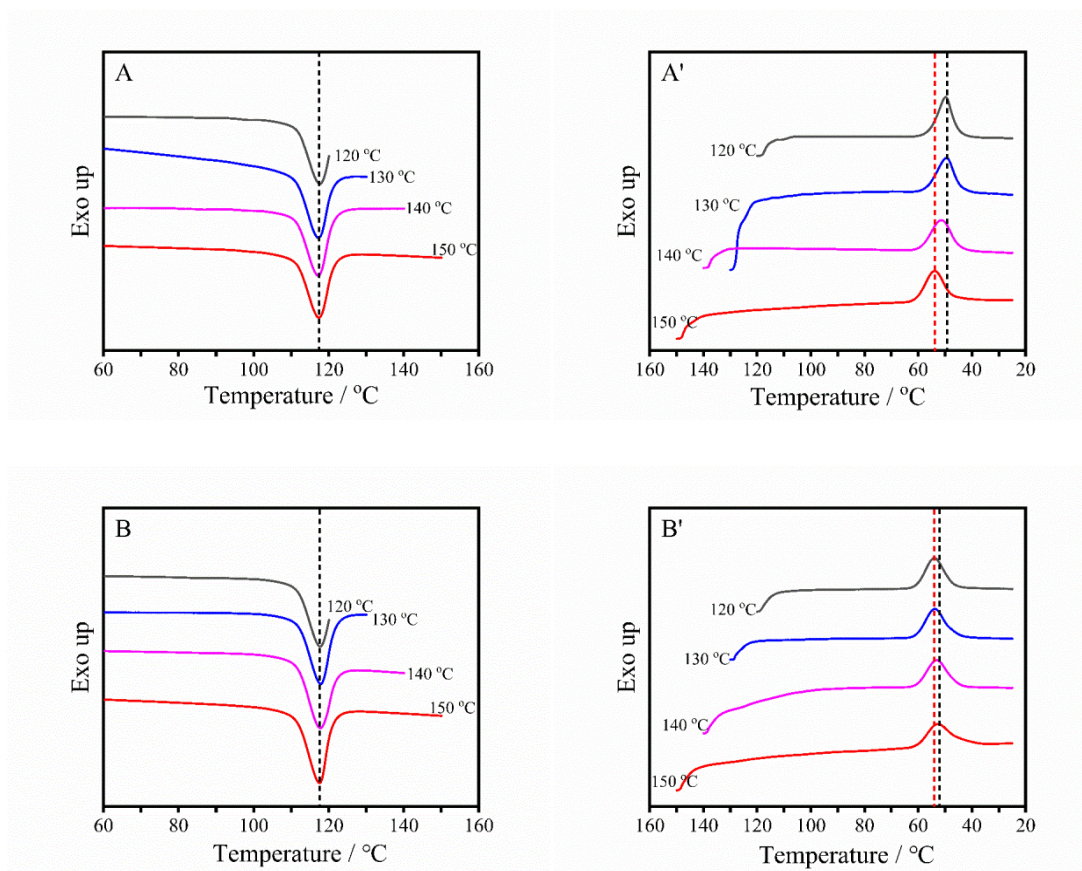

FIG. 7. Initial heating temperature related heating and cooling curves of the form I samples of PB-1 transformed from the form II obtained by heating to 120 °C (A and A′) and 150 °C (B and B′). (The thermal protocols are shown in Scheme 1C.)

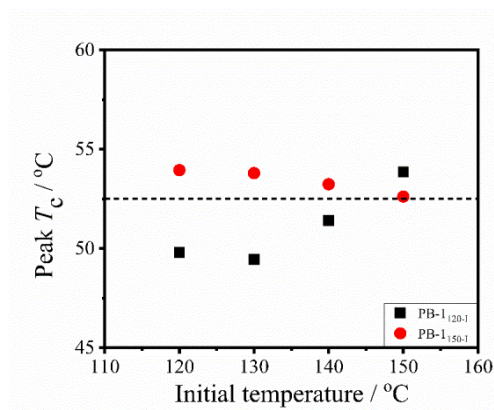

FIG. 8. Initial heating temperature related peak $T_c$s of form I samples of PB-1 formed after heating to 120 °C and 150 °C. (The thermal protocols are shown in Scheme 1C.)

Thermodynamically and kinetically controlled polymer crystallization is a typical non-equilibrium and complex process [20, 46, 47]. The conformational entropy of polymers depends on the chain conformations, which in turn depend on the temperature [51]. Adam and Gibbs established a link between relaxation times and temperature-dependent conformational entropy [30]. The Adam-Gibbs method presents an expression for the relaxation time $\tau$ of a polymer, which includes the conformational entropy ($S_c$) of the polymer melting at temperature $T$. The expression is

$$\tau = \tau_0 exp\left(\frac{C\Delta\mu}{TS_c(T)}\right) \quad (1)$$

where $\Delta\mu$ is the regular free energy barrier for rearrangements (for each molecule in the cooperative group), and C is a constant, $S_c(T)$ is defined as $S_c(T) = S_{melt} - S_{crystal}$, and $S_c(T)$ can be determined by the Adam-Gibbs method in the following equation

$$S_c(T) = \Delta S_{fus} - \int_T^{T_{fus}} C_p^{melt}(T') - C_p^{cryst}(T')\, d\log T' \quad (2)$$

where $T_{fus}$ is the temperature, $\Delta S_{fus} = \Delta H_{fus}/T_{fus}$ is the melting entropy Therefore, $S_c$ measures all of the melt entropy in addition to the vibration contribution ($S_{crystal}$). Eqs. (1) and (2) can be used to understand the typical nonequilibrium process during crystallization of polymers with inert chains crystallizing from melts at different temperatures. The illustration of the temperature related relaxation and entropy evolution of polymers via nonequilibrium process are shown in Fig. 9. As mentioned above, PB-1 is an inert polymer with long chain relaxation times. It thus tends to retain its previous structure and properties, including entropy, via nonequilibrium processes, as illustrated in Fig. 9.

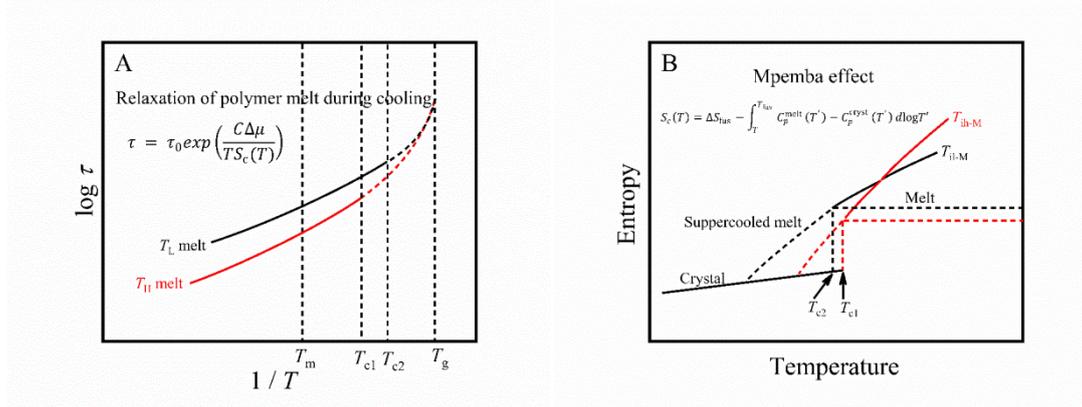

FIG. 9. Illustration of the temperature related relaxation time (A) and entropy (B) of PB-1 the samples cooled from low and high initial heating temperatures with nonequilibrium process.

To understand the phenomena in present work, we propose a possible explanation for the experimental results. The energy barrier for crystal formation [52, 53] and the polymer amorphous chain conformation [51] are dependent on temperature, where the energy barrier refers to initiation of polymer crystallization[52, 53]. When the temperature changes, the nucleation barrier or the activation barrier of the mesomorphic layer changes. Therefore, it is necessary to adjust or change the conformation of the nascent PB-1 chains in the amorphous phase during heating or cooling. The nascent PB-1 used in this experiment was synthesized at 35 $^{o}$C with low entropy of the amorphous melt. In general, polymers crystallizing from a lower initial melt temperature crystallize faster, because the chain conformation formed at the initial temperature is closer to the crystalline conformation. However, the nascent PB-1 shows an opposite trend after heating to various initial temperatures. This should be related to the chain dynamics of PB-1. Previous studies have reported that the radius of gyration of polymer chains does not change much during crystallization, indicating that adjacent chain segments adopt the correct extended conformation for the crystal lattice and crystallize in situ [54, 55]. During holding at the initial heating temperatures, PB-1 may form mesophase or local structure [25]. Since the chain activity increases with temperature, the mesophase or local structure will grow to a larger size at higher temperatures. In the subsequent crystallization process, PB-1 chains within the larger mesophase or local structure may be able to more readily crystallize via nonequilibrium dynamics. The larger mesophase might enable the more efficient ordering of the PB-1 chains [56, 57]. On the contrary, PB-1 forms mesophases or local structures of smaller size at lower temperatures, which would have constraining effect on the chain dynamics compared to larger size structures. The smaller size would reduce the dynamics of the PB-1 chains, limit the diffusion and sliding of the chains, and further constrain the PB-1 chain conformation during crystallization [58, 59]. In other words, the polymers have a more flexible chain conformation in the melt at high temperature, and the conformation is more readily changed during cooling. PB-1 crystallization after holding at low temperature is slow, contrary to the general situation, i.e. PB-1 crystallization shows the Mpemba effect. Due to the long relaxation time of the slow PB-1 chain, the conformational entropy change of the nascent PB-1 lags behind the temperature change when the initial heating temperature is below the equilibrium melting point of form I [50]. According to the proposed phase diagram of PB-1 conformational propensities [60], the conformational propensity for the 11/3 helix (characteristic of form II) with high entropy would increase with temperature when the PB-1 sample is heated to a temperature below the equilibrium melting point of form I. However, above the equilibrium melting point of form I, states favoring 3/1 helices would be disfavored compared to 11/3 helices due to relaxation, and the conformational propensities for 11/3 helices would instead become those for 3/1 helices when the temperature is lower than 70 $^{\circ}$C. The higher the heating temperature, the longer transition time of PB-1 would be [60, 61], sinceit is related to the formation of form I of PB-1. The observed Mpemba effect in PB-1 crystallization is related to the formation of form II which is determined by the initial heating temperature corresponding to the conformational propensities for 3/1 helices or 11/3 helices.

As shown the illustration in Fig 9, the relaxation and entropy of the polymer chains depend on temperature and the time related to the temperature. In order to understand the crystallization behavior and the Mpemba effect observed in the nascent PB-1 samples, the nascent PB-1 samples were heated and cooled at different heating and cooling rates, or kept at 130 ℃ for different times and then isothermally crystallized at 70 ℃. The obtained results are shown in Fig. 10. The results in Fig. 10A and 10B display that the heating rate and cooling rate cannot affect the Mpemba effect in the melt crystallization of the nascent PB-1. However, the $T_c$ is affected more significantly by the cooling rate than by the heating rate. The results in Fig. 10C and 10D show that $T_c$ increases and $t_{1/2}$ decreases with the holding time at the initial heating temperature, and the Mpemba effect in the crystallization of nascent PB-1 is observed with the reduced heating or cooling rate even though the effect was diminished, which indicate that the conformational propensities for 11/3 helices increase (or 3/1 helices decrease) with temperature and relaxation time. It is known that the form II is a crystalline form with 11/3 helices packed in tetragonal cells and form I is a crystalline form with 3/1 helices packed in hexagonal cells [62, 63]. The conformational propensities for 11/3 helices would enhance the crystallization of form II. Therefore, we calculated the crystallinity of form II by further heating the samples used in Fig. 4 and Fig. 7 and the obtained results are shown in Fig. 11. The results in Fig. 11 demonstrate the crystallinity of form II obtained after the further heating the form II and form I, where the crystallinity corresponding to the form II and form I obtained after cooling from 150 ℃ is greater than that of form II and form I obtained after cooling from 120 ℃, and the crystallinity corresponding to form I is greater than that of the corresponding form II. The Mpemba effect in the crystallization of PB-1 is due to the content of form II and its 11/3 helical precursor state in the PB-1 samples.

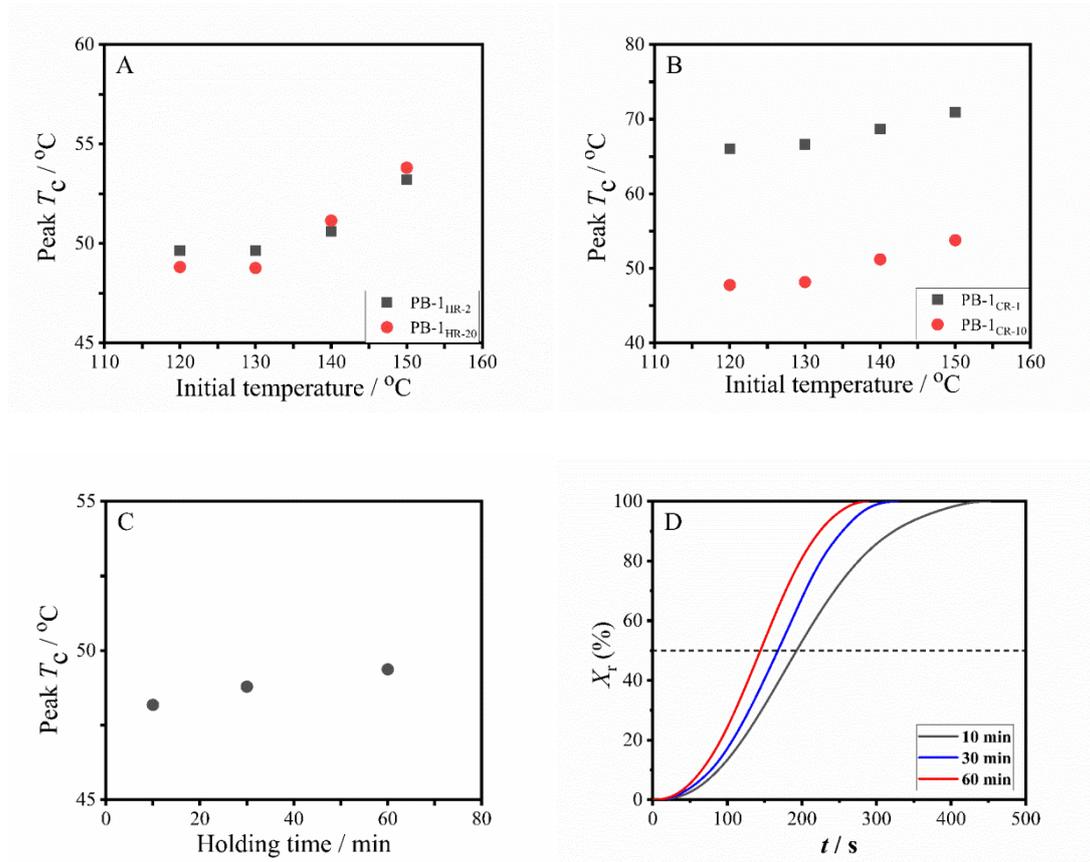

FIG. 10. Initial heating temperature related peak $T_c$s of nascent PB-1 form II samples with the indicated rates of heating (A), cooling (B) and holding for various time when initially heating to 130 °C (C). The crystallinity evolution of the nascent PB-1 during isothermal crystallization at 70 °C after heating to 130 °C and holding for various time (D). (The thermal protocols are shown in Scheme 1D.)

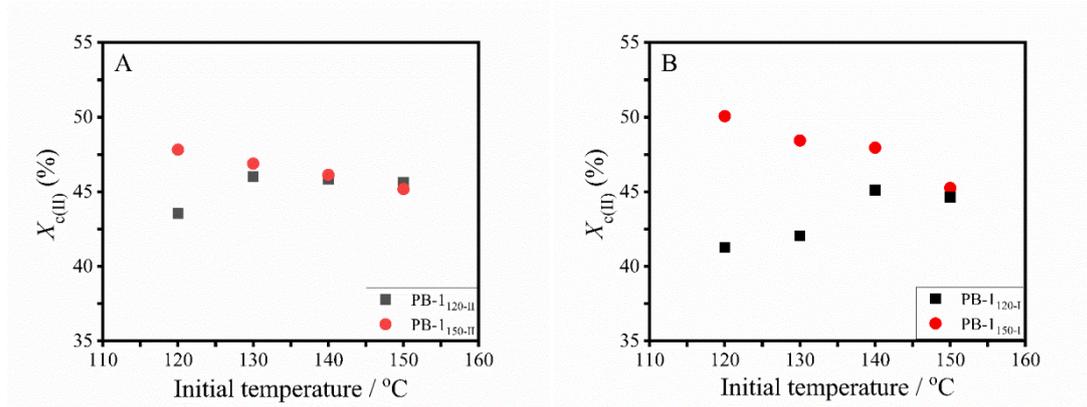

FIG. 11. Initial heating temperature related crystallinities of form II obtained from the form II and form I samples of PB-1 originally formed by heating to 120 $^{\circ}$C and 150 $^{\circ}$C.

The crystallization temperature difference ($T_i - T_c$) of PB-1 form II and form I between the initial heating temperature and the peak of $T_c$ is plotted as a function of the cooling start temperature in Fig. 12. The intersection of the two lines related to the ($T_i - T_c$)s of form II and form I obtained after heating to 120 $^{\circ}$C and 150 $^{\circ}$C corresponds to the equilibrium melting point of PB-1 form I. In other words, the entropy of the PB-1 melt cannot easily reach a maximum when the initial heating temperature is lower than the equilibrium $T_m$ of form I; the Mpemba effect cannot easily be observed when the initial heating temperature is higher than the equilibrium $T_m$ of form I.

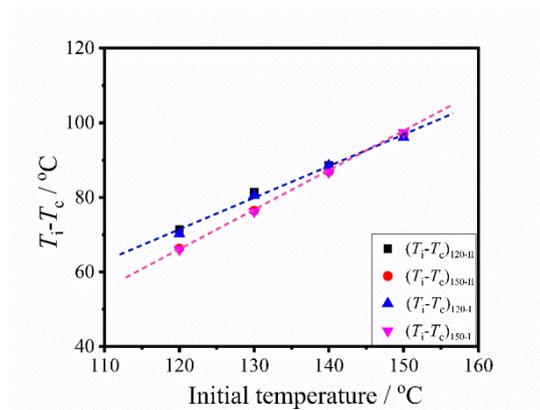

FIG. 12. The difference between the $T_i$ and $T_c$ as a function of initial temperature for crystallization.

## 4. Conclusion

In summary, the Mpemba effect was observed in the melt crystallization process of PB-1 under different heat treatment conditions, which is the first time that this effect has been reported in the melt crystallization process of polymers. The Mpemba effect in the crystallization behavior is attributed to changes in the conformation of the PB-1 chain associated with the equilibrium melting point of PB-1 form I during crystallization. A high initial temperature below the equilibrium melting point of form I favors the formation of the 11/3 helix corresponding to form II, which benefits the crystallization of form II and leads to the Mpemba effect. The reported observations shed light on the crystallization and transformation process of PB-1 and contribute to the understanding of the polymer dynamics.

**Notes**

The authors declare no competing financial interest.

**ACKNOWLEDGEMENTS**

The authors acknowledge NSFC for financial support. This work was carried out with the support of 1W2A beamline at Beijing Synchrotron Radiation Facility (BSRF, λ = 1.54 Å) and BL16B1 of Shanghai Synchrotron Radiation Facility (SSRF, λ=1.24 Å).

**Data Availability Statement**

Data openly available in a public repository.

The data that support the findings of this study are openly available in [repository name] at [URL].